\documentclass[prd, twocolumn, nofootinbib]{revtex4}

\usepackage{epsfig} 
\usepackage{url} 

\newcommand{\beq}{\begin{equation}}
\newcommand{\eeq}{\end{equation}}
\newcommand{\beqa}{\begin{eqnarray}}
\newcommand{\eeqa}{\end{eqnarray}}

\def\fun#1#2{\lower3.6pt\vbox{\baselineskip0pt\lineskip.9pt
  \ialign{$\mathsurround=0pt#1\hfil##\hfil$\crcr#2\crcr\sim\crcr}}}

\newcounter{tempref}

\begin{document} 

\title{Resource Letter: %DEAU-1: 
Dark Energy and the Accelerating Universe} 
\author{Eric V.\ Linder} 
\affiliation{Berkeley Lab, University of California, Berkeley, CA 94720} 

\date{\today} 

\begin{abstract} 
This Resource Letter provides a guide to the literature on dark energy 
and the accelerating universe.  It is intended to be of use to researchers, 
teachers, and students at several levels.  Journal articles, books, and 
websites are cited for the following topics: Einstein's cosmological 
constant, quintessence or dynamical scalar fields, modified cosmic gravity, 
relations to high energy physics, cosmological probes and observations, 
terrestrial probes, calculational tools and parameter estimation, teaching 
strategies and educational resources, and the fate of the universe. 
\end{abstract} 

\maketitle

\section{Introduction} \label{sec:intro} 

Acceleration of the expansion of the universe is one of the most 
exciting and significant discoveries in physics, with implications 
that could revolutionize theories of quantum physics, gravitation, 
and cosmology.  With its revelation that close to the three-quarters 
of the energy density of the universe, given the name dark energy, 
is of a new, unknown origin and that its exotic gravitational 
``repulsion'' will govern the fate of the universe, dark energy and 
the accelerating universe becomes a topic not just of great interest 
to research physicists but to science students at all levels. 

This Resource Letter endeavors to guide teachers and interested 
scientists through the yet-inchoate field of dark energy. 
This subject is continuously developing and is somewhat amorphous, not 
well bounded due to our ignorance of from what part of physics the 
explanation of cosmic acceleration will stem.  While this raises 
challenges for pedagogy, it also offers opportunities for giving a 
broad overview of the interplay of fundamental physics, cosmology, 
and astrophysics and a living example of how physics is a dynamic 
field with ongoing discoveries. 

The emphasis here is on a broad review along the lines of important 
concepts rather than specific theories, which rise and fall in 
popularity, and on the interaction of concepts with observational 
data.  Likewise, the focus for the resources listed is on the widest 
pedagogical usefulness rather than necessarily the original or 
latest literature; 
arXiv.org (Ref.~\ref{arxiv}), the NASA Astrophysical Data System Abstract 
Service (Ref.~\ref{ADS}), and SPIRES high energy physics (Ref.~\ref{spires}) 
online servers have efficient searches for the 
specific literature.  The references given to major conferences and 
to websites supplement these with overviews of the latest results and 
trends in thought. 

Historically, several issues led to dissatisfaction with the idea of 
a universe presently dominated by matter.  These included the age 
problem of old objects in what dynamics would predict to be a young 
universe, the pattern of the large scale clustering of galaxies and 
clusters of galaxies, and the stability problem of a universe with 
density different from the critical density (or nonzero spatial 
curvature).  While there was a puzzle on the fundamental physics 
side -- why did Einstein's cosmological constant vanish -- this was 
predominantly happily swept out of sight by an assumption that some 
symmetry would eventually explain it. 

Observational evidence for a new component of the universe that brought 
relief to the astrophysical issues swept the scientific community in 
1998, from two independent groups measuring the cosmological 
distance-redshift relation of Type Ia supernovae, with quick support 
in the following years from cosmic microwave background data consistent 
with a critical density universe and more detailed large scale structure 
observations limiting the matter component contribution.  However, the 
consistent picture on the cosmological side brought to the forefront the 
puzzle on the quantum side: from where did a cosmological constant of 
the precise magnitude needed arise, and was it in fact a true, constant 
vacuum energy? 

The new picture of the universe was a dramatic Copernican revolution. 
From abandonment of humans possessing a special spatial location in the 
universe, to abandonment of baryon chauvinism -- that we are made of 
the typical stuff of the universe -- due to dark matter, we now must 
abandon ``rest mass chauvinism'' -- the large majority of the universe 
is not made of components dominated by their rest mass, like most of what 
we know, but rather is 
energy-like.  Moreover, this dark energy violates the strong energy 
condition, having effectively negative gravitating mass, so that we 
are forced to abandon Newton and ask what happens when the force of 
gravity is no longer attractive. 

These are extraordinary, dramatic questions to face.  The answers, 
which we in no wise know yet, will rewrite textbooks.  In \S\ref{sec:big} 
we present resources giving a big picture overview of the impact that the 
discovery of the accelerating universe has had on cosmology research. 
Non-cosmological approaches to dark energy are briefly discussed in 
\S\ref{sec:lab}.  Classes of possible solutions to the dark energy puzzle 
are given thumbnail descriptions in \S\ref{sec:models}, and observational 
approaches to detecting and characterizing dark energy in \S\ref{sec:obs}. 
In \S\ref{sec:tools} we list resources for some standard calculational methods 
and tools.  \S\ref{sec:teach} discusses approaches to teaching dark energy 
at various levels, and \S\ref{sec:fate} goes into more detail concerning 
the fascinating and complex topic of life in an accelerating cosmos and 
the fate of the universe.  \S\ref{sec:review} presents a compilation of 
review articles suitable for broad exploration of the field or 
aspects of the field.  References denoted (E) are at a popular science 
level, while those listed as (I) are for readers with some familiarity 
with the basics of cosmology or particle physics, and advanced (A) material 
requires a technical background in general relativity, field theory, or 
detailed cosmology.  Further references in the form of books, conferences, 
and research and education websites appear in \S\ref{sec:other}.

\section{Dark Energy in Cosmology \label{sec:big}}

With the discovery of the accelerated expansion of the universe in 1998, 
and its corroboration and more detailed measurement soon thereafter, 
cosmology and particle physics research took stock of the implications 
of these extraordinary results and shifted into high gear for exploration 
of the dark universe.  For a comprehensive view of astrophysics in 
general just before this explosion of effort, see Ref.~\ref{decadal}, 
with a partial update in Ref.~\ref{decadalmid}. 
A study by a National Academy of Sciences panel tying together the 
possible revolutions in cosmology and quantum physics led to  
Refs.~\ref{q2c}-\ref{q2cimp}.  A recent European overview, from a more 
astrophysical perspective, is Ref.~\ref{esa}, while Ref.~\ref{detf} 
from a DOE-NASA-NSF committee presents a more technically focused report 
on dark energy. 

\vspace{0.1in}

\noindent{\bf Research Literature Searches} 

\begin{enumerate}

\item \label{arxiv} 
arXiv.org preprint server \url{http://arXiv.org}.  
Up to the minute collection of research articles, together with a good 
search facility. 
Dark energy related work appears most frequently in astro-ph, gr-qc, 
hep-th, and hep-ph categories. 

\item \label{ADS} 
Harvard-Smithsonian 
Center for Astrophysics/NASA Astrophysics Data System abstract service 
\url{http://adsabs.harvard.edu/ads_abstracts.html}.  
Comprehensive collection of research article titles and abstracts, with 
links to published versions, or unpublished versions on Ref.~\ref{arxiv}. 
Good search facility. 

\item \label{spires}
SPIRES high energy physics literature database 
\url{http://www.slac.stanford.edu/spires}. 
Database of research articles, with search facility and citation 
statistics.  Oriented toward high energy physics but includes most 
cosmology journals. 

\setcounter{tempref}{\theenumi}

\end{enumerate} 

\vspace{0.1in}

\noindent{\bf Committee Overviews}

\begin{enumerate}

\setcounter{enumi}{\thetempref}

\item \label{decadal} 
{\bf Astronomy and Astrophysics in the New Millennium} (2001) 
\url{http://www7.nationalacademies.org/bpa/aanm.html}.  Latest in series 
of decadal surveys of the state of astronomy and astrophysics.  Because of 
the timing (produced by 1999), contains little on dark energy but provides 
a very broad overview of astrophysics at that time. (I)

\item \label{q2c} 
{\bf Connecting Quarks with the Cosmos: Eleven Science Questions for the 
New Century} (2003) 
\url{http://www.nap.edu/catalog.php?record_id=10079}.  Influential 
assessment of key physics questions unifying particle physics and 
cosmology.  Dark energy is one of the eleven questions, plus several of 
the others may be related to the accelerating universe.  Gives both 
scientific background as well as programmatic views. (I) 

\item \label{q2cimp} 
{\bf A 21st Century Frontier of Discovery: The Physics of the Universe} (2004) 
\url{http://www.ostp.gov/html/physicsoftheuniverse2.pdf}.  Report from 
the National Science and Technology Council discussing strategic 
implementation of the recommendations of Ref.~\ref{q2c}, including dark 
energy experiments. (I)

\item \label{decadalmid} 
{\bf Review of Progress in Astronomy and Astrophysics Toward the Decadal 
Vision} (2005) 
\url{http://www7.nationalacademies.org/bpa/Mid_Course_Review_Home.html}.  
Prompted by increased interaction between particle physics and cosmology, 
in particular Ref.~\ref{q2c}, this reviewed progress since Ref.~\ref{decadal}. 
Contains some mention of dark energy, mostly from the programmatic point 
of view. (I)

\item \label{detf} 
{\bf Report of the Dark Energy Task Force} (2006) 
\url{http://arxiv.org/abs/astro-ph/0609591}.  From an advisory 
subcommittee for DOE, NASA, and NSF.  More technical assessment 
of experimental methods for probing dark energy, with particular attention 
to practical issues, ground vs.\ space observations, and a series of 
experimental stages.  Contains some background material and a long 
technical appendix. (A) 

\item \label{esa} 
{\bf Report by the ESA-ESO Working Group on Fundamental Cosmology} (2006)
\url{http://arxiv.org/abs/astro-ph/0610906}.  European assessment of 
cosmology issues, similar to Refs.~\ref{q2c}, \ref{q2cimp}, but with less 
emphasis on particle physics and more on astrophysics.  Good overview 
sections with several discussions of dark energy and the accelerating 
universe. (I) 

\setcounter{tempref}{\theenumi}

\end{enumerate}

\section{Dark Energy in the Lab \label{sec:lab}}

Before investigating the main, cosmological aspects of the accelerating 
universe, 
we briefly mention the more speculative approaches to exploring a modified 
gravity origin of cosmic acceleration through laboratory and solar system 
tests.  If gravity is altered due to extra dimensional effects, the 
excitation (Kaluza-Klein) modes of the extra dimensions might be detectable 
as particles in high energy accelerators; Ref.~\ref{ilc} considers the 
possibility of next generation particle accelerators for probing dark 
energy.  The effects of cosmic acceleration could arise from weakening 
the force of gravity relative to the inverse square law at long distances; 
the modification of the structure of gravity could 
also show up at micron scales (e.g.\ again due to extra dimensional effects) 
or scales related to the spacetime curvature (solar system scales for 
orbits bound by the mass of the Sun).  Recent results from laboratory 
tests of the gravitational inverse square law appear in Ref.~\ref{adel}, 
while Refs.~\ref{llr}-\ref{llr2} present lunar laser ranging constraints, and 
Ref.~\ref{mars} discusses the use of planetary orbits.  The great 
preponderance of possible origins for dark energy, including many 
modifications of gravity, are only detectable through cosmological 
signatures, however.  Ref.~\ref{easyway} gives a simple discussion of why 
some ``direct'' detection techniques are not generally expected to be 
fruitful approaches.

\begin{enumerate} 

\setcounter{enumi}{\thetempref}

\item \label{ilc} 
Linear Collider Connections to Astrophysics and Cosmology 
\url{http://www.physics.syr.edu/~trodden/lc-cosmology}.  Working group 
of the American Linear Collider Physics Group for the International 
Linear Collider (ILC); no report issued yet, check website for future 
documents. 

\item \label{adel} 
``Tests of the Gravitational Inverse-Square Law below the Dark-Energy 
Length Scale,'' D.J.\ Kapner, T.S.\ Cook, E.G.\ Adelberger, J.H.\ Grundlach, 
B.R.\ Heckel, C.D.\ Hoyle, and H.E.\ Swanson, Phys.\ Rev.\ Lett.\ {\bf 98}, 
021101 (2007). Latest results from laboratory experiments probing 
submillimeter gravity. (A) 

\item \label{llr} 
``Testing Gravity via Next-Generation Lunar Laser-Ranging,'' T.W.\ Murphy, 
Jr., E.G.\ Adelberger, J.D.\ Strasburg, C.W.\ Stubbs, and K.\ Nordtvedt, 
Nucl.\ Phys.\ B (Proc.\ Suppl.) {\bf 134}, 155-162 (2004).  Motivation 
and next generation techniques for using lunar laser ranging to probe 
gravity. (I) 

\item \label{llr2} 
APOLLO Project: \url{http://physics.ucsd.edu/~tmurphy/apollo}.  Ongoing 
experiment to improve limits on the Equivalence Principle and gravitational 
properties through lunar laser ranging.  Contains physics background 
materials and technical description. (I)

\item \label{mars} 
``The Accelerated Universe and the Moon,'' G.\ Dvali, A.\ Gruzinov, and 
M.\ Zaldarriaga, Phys.\ Rev.\ D {\bf 68}, 024012 (2003).  Theoretical 
discussion of how modifying the structure of gravity to lead to cosmic 
acceleration could influence orbits, such as in the solar system. (A)

\item \label{easyway} 
``Dark energy the easy way?,'' 
\url{http://supernova.lbl.gov/~evlinder/easyde.html}. Brief discussion 
of various direct methods for seeing dark energy and why they are not 
so easy. (E)

\setcounter{tempref}{\theenumi}

\end{enumerate}

\section{Theoretical Approaches to Dark Energy \label{sec:models}}

This section aims to give a flavor of classes of models proposed for 
dark energy, without going into any detail on specific models or 
attempting to be comprehensive.  

Einstein proposed a {\it cosmological constant\/} to counterbalance 
gravitational 
attraction; its Lorentz invariant form leads to another interpretation 
in terms of ground state, or zeropoint, vacuum energy of quantum fields. 
The energy density of a cosmological constant is unchanging with the 
cosmic expansion, and its pressure to energy density, or equation of state, 
ratio is fixed at $w\equiv p/\rho=-1$.  Scalar fields that do not sit 
at the minimum of their potential are dynamical, called {\it quintessence\/} 
when they have canonical kinetic terms and minimal coupling to gravity; 
they possess a time varying energy density and generally a time varying 
equation of state.  {\it Coupled dark energy\/} usually involves a scalar 
field with coupling to dark matter; {\it k-essence\/} involves a scalar 
field with a noncanonical kinetic term.  

Other models include {\it unified dark energy\/} 
models, also called Chaplygin gas, quartessence, or mocker models, 
attempting to explain dark energy and dark matter from a single physical 
origin.  {\it Holographic dark energy\/} tries to explain vacuum energy 
through a summation over modes of zeropoint energy limited by a 
holographic conception of number of degrees of freedom.  {\it Backreaction 
models\/} attempt to obtain merely the perception of acceleration through 
effects of nonlinear matter structures on the cosmic expansion, or 
specialized distributions of matter such as Lema{\^\i}tre-Tolman-Bondi 
models.  Modified 
gravity origins for cosmic acceleration include {\it scalar-tensor 
theories\/} of a scalar field nonminimally coupled to spacetime curvature 
and $f(R)$ {\it theories\/} with more complicated functions of the Ricci 
spacetime curvature than the Einstein-Hilbert action, and there exist 
more complicated functions of the full curvature tensor as well, plus 
extradimensional theories such as {\it braneworld cosmology\/}.  As 
stated, this is a partial sampling of the many ideas attempting to 
explain the origin of cosmic acceleration.  See the review articles cited 
in \S\ref{sec:review} for further discussion and use Ref.~\ref{arxiv} to 
search on the terms in detail.

\section{Observational Approaches to Dark Energy \label{sec:obs}}

Here we summarize observational methods for probing dark energy and 
the accelerating universe, again without going into detail on technical 
specifics or proposed programs.  While direct detection of acceleration 
is unlikely (see Ref.~\ref{easyway}), measurement of the cosmic expansion 
through geometric quantities such as distances depending only on the 
cosmic scale factor and spatial curvature is feasible.  The slowing and 
acceleration of scales with time, $a(t)$, map out the cosmic environment 
history like the lesser and greater growth of tree rings map out the 
Earth's climate history.  The technique that discovered the acceleration, 
the Type Ia supernova magnitude-redshift test, is such a geometric test.  
Once the supernovae are calibrated through measuring their properties, 
the magnitude, or received flux, 
measures the distance through the cosmological inverse square law, giving 
the lookback time to the explosion, while the measured redshift gives 
the scale factor; hence we obtain $a(t)$. 

Similar geometric probes can be gotten from distance ratios arising from 
gravitational lensing observations (both weak lensing where one statistically 
measures subtle shape distortions in distant galaxies by foreground mass, 
and strong lensing where multiple images occur) baryon acoustic 
oscillations (statistically measuring the scale of preferred spatial 
clustering of galaxies or other baryon-rich objects relative to the 
primordial sound horizon scale from the cosmic microwave background 
recombination epoch), and possibly other techniques. 

As a further consequence of the accelerated expansion,  
the growth of large scale structure such as galaxies and clusters 
of galaxies is suppressed.  Structure grows through gravitational 
instability, but the stretching of space works against aggregation of 
mass just as a person finds it hard to join a large group of friends 
if they are at the end of an escalator running the wrong way.  Observations 
of the growth of massive structures through weak lensing, or the formation 
and evolution of galaxies and clusters, e.g.\ cluster 
counts (the abundance of clusters in a certain mass range), employing large 
surveys in the optical, submillimeter (Sunyaev-Zel'dovich effect), or 
X-ray bands, or again using weak gravitational lensing to detect mass, can 
thus give more indirect evidence of acceleration.  Other than for 
weak lensing, which directly measures mass, these techniques also need 
to translate the observed light to the true mass.  Methods that incorporate 
mass or mass and gas measurements in addition to simple geometry must 
separate out the astrophysics of galaxy and cluster formation and evolution 
to extract the cosmological dark energy characteristics. 

The cosmic microwave background (CMB) radiation is a precise probe of 
many cosmological quantities, but is not particularly directly incisive 
on dark energy, at least in present usage.  Basically the CMB gives a 
snapshot of the early universe, at 0.003\% of its present age, and so 
does not reveal detailed characteristics 
of the late time acceleration of the universe. 

These observational approaches to dark energy are discussed from both 
scientific and proposed experimental program points of view in the 
Committee Overviews 
cited in \S\ref{sec:big} and in the review articles in \S\ref{sec:review}.

\section{Tools for Dark Energy Cosmology \label{sec:tools}} 

Analysis tools to compare cosmological models with observations or 
simulations, while not specific to dark energy, are an important ingredient 
of understanding the accelerating universe and the physics behind it.  
These tools range from calculators of distances or other cosmological 
relations to solvers of coupled perturbation growth equations to fitters 
of cosmological parameters to various data sets.  Cosmology fitters use 
various techniques to calculate the likelihoods of models, from chi-square 
minimization to the Fisher information matrix (Gaussian approximation of 
the likelihood surface near the maximum) to Monte Carlo techniques. 
We present here a selection of general discussions and publicly available 
codes to carry out these tasks.

\begin{enumerate} 

\setcounter{enumi}{\thetempref}

\item \label{tegtayhea}
``Karhunen-Lo{\`e}ve Eigenvalue Problems in Cosmology: How Should We 
Tackle Large Data Sets?,'' M.\ Tegmark, A.N.\ Taylor, A.F.\ Heavens, 
Ap.\ J.\ {\bf 480}, 22-35 (1997). Technical and mathematical basis 
for Fisher information matrix analysis. (A) 
%http://arxiv.org/abs/astro-ph/9603021

\item \label{eishuteg}
``Cosmic Complementarity: Probing the Acceleration of the Universe,'' 
M.\ Tegmark, D.J.\ Eisenstein, W.\ Hu, \& R.G.\ Kron, arXiv:astro-ph/9805117. 
More accessible guide to Fisher information matrix analysis than 
Ref.~\ref{tegtayhea}, with practical examples. (A) 
%``Cosmic Complementarity: Joint Parameter Estimation from CMB Experiments 
%and Redshift Surveys,'' D.J.\ Eisenstein, W.\ Hu, \& M.\ Tegmark, Ap.\ J.\ 
%{\bf 518), 2-23 (1999). (A) 
% http://arxiv.org/abs/astro-ph/9807130

\item \label{doomsdayapx}
``Observational Bounds on Cosmic Doomsday,'' R.\ Kallosh, J.\ Kratochvil, 
A.\ Linde, E.V.\ Linder, M.\ Shmakova, JCAP {\bf 0310}, 015 (2003).  The 
appendix gives a simplified, step by step guide to applying the Fisher 
information matrix. (I) 
%http://arxiv.org/abs/astro-ph/0307185

\item \label{nijfis}
``Design a Dark Energy Experiment,'' E.V.\ Linder, 
\url{http://supernova.lbl.gov/~evlinder/design.pdf}.  Summer school lecture 
on relating data to theory, with emphasis on systematic uncertainties. (I) 

\item \label{minuit}
Minuit \url{http://wwwasdoc.web.cern.ch/wwwasdoc/minuit/minmain.html}. 
CERN library program for function minimization (e.g.\ chi-squared) and 
error analysis. 

\item \label{cosmomc}
CosmoMC \url{http://cosmologist.info/cosmomc}.  Markov chain Monte Carlo 
program for cosmological parameter estimation. 

\item \label{cosmonest}
CosmoNest \url{http://www.cosmonest.org}.  Implements nested sampling 
for calculation of Bayesian evidence, or cosmological model likelihoods. 

\item \label{cmbfast}
CMBfast \url{http://cfa-www.harvard.edu/~mzaldarr/CMBFAST/cmbfast.html}. 
Program for calculating cosmic microwave background temperature and 
polarization anisotropies and matter power spectrum.  Standard version 
has constant equation of state for dark energy (or table of values). 

\item \label{camb}
CAMB (Code for Anisotropies in the Microwave Background) 
\url{http://camb.info}.  Extension of CMBfast with high accuracy matter 
power spectra and Monte Carlo interface.  Standard version has constant 
equation of state for dark energy. 

\item \label{cmbeasy}
CMBeasy \url{http://www.cmbeasy.org}.  Extension of CMBfast with numerous 
dark energy parameterizations and Markov Chain Monte Carlo generation. 
Has extensive documentation and graphical user interface. 

\item \label{cmbcross} 
Cross\_CMBfast \url{http://www.astro.columbia.edu/~pierste/ISWcode.html}. 
Extension of CMBfast for computing integrated Sachs-Wolfe correlations. 
Includes $w_0$-$w_a$ dark energy model and constant sound speed. 

\item \label{wrightcalc}
Cosmology Calculator 
\url{http://www.astro.ucla.edu/~wright/ACC.html}.  Javascript to compute 
cosmological distances and ages. 

\item \label{frenchfit}
Kosmoshow \url{http://marwww.in2p3.fr/renoir/Kosmoshow.html}.  Cosmology 
fitter to supernova data using chi-squared minimization.  Allows extensive 
control of data and error properties, and includes plotting routines. 
Graphical user interface. 

\item \label{conleyfit}
Simple Cosfitter \url{http://qold.astro.utoronto.ca/conley/simple_cosfitter}. 
Cosmology fitter concentrating on supernovae magnitude-redshift data. 
Includes terms for supernova color and stretch parameters. 

\item \label{tonryfit}
Supernovae chi-squared evaluator 
\url{http://www.ifa.hawaii.edu/~jt/SOFT/snchi.html}.  Calculates chi-squared 
based on the distance-redshift relation. 

\item \label{snoc}
SNOC (Supernova Observation Calculator) 
\url{http://www.physto.se/~ariel/snoc}.  Monte Carlo simulator of 
supernovae data and cosmology parameter fitter.  Contains numerous 
astrophysical effects such as gravitational lensing and dust extinction 
and is useful for systematic uncertainty studies. 

\item \label{detffast}
DETFast \url{http://www.physics.ucdavis.edu/DETFast}.  
Java applet producing cosmological parameter Fisher matrix likelihood 
contours for various simulated data sets used by the Dark Energy Task 
Force (see Ref.~\ref{detf}). 

\setcounter{tempref}{\theenumi}

\end{enumerate}

\section{Approaches to Teaching Dark Energy \label{sec:teach}}

Teaching and communication of aspects of dark energy and the accelerating 
universe can take place at many levels, as seen for example in 
the Review Articles of \S\ref{sec:review} and Education Websites of 
\S\ref{sec:edu}.  On a formal level, students 
should certainly be familiar with the expansion of the universe and 
the behavior of matter and radiation equation of states before being 
introduced to the acceleration of the expansion.  Students with exposure 
to classical mechanics can appreciate the simple scale field 
Lagrangian, equation of state ratio, and Klein-Gordon equation.  A 
phenomenological approach to modifying the Friedmann expansion equation 
can create understanding of the effects of a five dimensional braneworld 
``leaking gravity'' model or string theory inspired barotropic terms 
(additional terms nonlinear in the matter density), especially 
when considering asymptotic future behavior.  

\begin{figure} 
\begin{center} 
\psfig{file=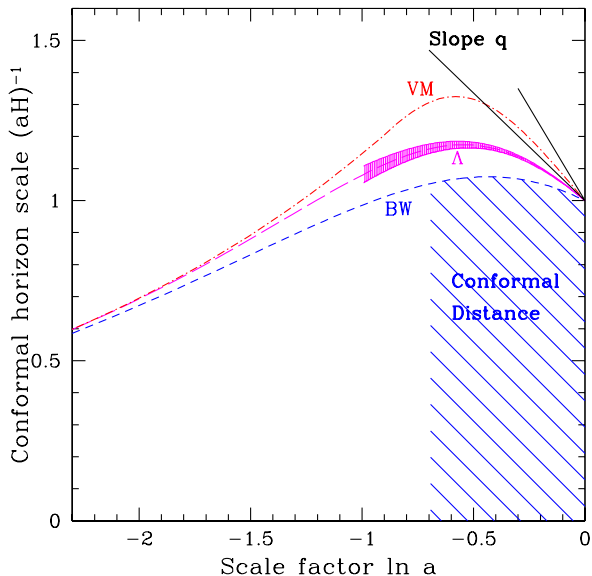,width=3.2in}
\caption{Plotting the conformal Hubble scale vs.\ the logarithmic 
scale factor one can readily visualize acceleration of the cosmic 
expansion through the slope of a cosmological model curve and distances 
through the area under the curve.  The dense shading around the 
cosmological constant $\Lambda$ curve represents a goal for 95\% confidence 
level constraints from next generation measurements, mapping both 
accelerating and decelerating phases and distinguishing between dark 
energy physics models. 
}
\label{fig:ahvisual}
\end{center}
\end{figure}

To visualize the geometric effects of the accelerating expansion, and 
to understand how the differences in dark energy models show up in 
cosmological observations, it is convenient to look at a conformal 
horizon diagram.  This allows students to see by eye concepts that are more 
difficult to pick out on a standard scale factor vs.\ time, $a(t)$, 
or distance-redshift $d(z)$ Hubble diagram.  In Figure~\ref{fig:ahvisual} 
comoving length scales would simply be horizontal lines and positive 
slopes for the expansion history curves correspond to decelerating epochs, 
and negative slopes are the sign of acceleration.  The slope of a curve 
at present is precisely the deceleration parameter today, $q_0$. 
The area under a curve is simply the conformal distance 
$\eta=\int d\ln a (aH)^{-1}$, precisely the quantity that enters in 
luminosity distances and angular distances.  Thus one can immediately 
see that distances in an 
accelerating universe are greater than they would be in a decelerating 
universe (with the same Hubble constant), or a less accelerating universe. 
One can even obtain the total equation of state of the universe $w_{\rm tot}$ 
and its running $w'=dw/d\ln a$ from the slope and curvature of the curves. 
These relations, and the clear application to inflation as well, make such 
a figure a useful and visual pedagogical tool.

For more advanced courses, 
scaling and tracking behavior, numerical solutions of the Klein-Gordon 
equation (or continuity equation with coupling between dark energy and 
matter), linear perturbation growth, effects on the distance-redshift 
Hubble diagram or cosmic microwave background temperature power spectrum 
are all accessible exercises to give insight into dark energy cosmology. 

Introductory level courses can go a considerable way into the main 
concepts with very modest amounts of math.  One key approach is emphasizing 
that Einstein's Equivalence Principle and the equivalence of energy and 
mass (``$E=mc^2$'') yield a picture of dark energy opposed to attractive 
gravity.  Einstein lessoned us that when determining an object's 
gravitational mass we must add up all the forms of equivalent energy, 
including pressure.  Hence students can be guided to the idea that an 
object with positive energy density, but sufficiently negative pressure, 
can effectively have a negative gravitational mass, and hence acts in 
repulsion not attraction.  Using the rubber sheet image of spacetime, 
while a ball of matter would cause a depression, experienced by other 
particles as gravitational attraction, a ball of dark energy causes a hill, 
perceived by other particles as gravitational repulsion (if not quite in a 
formal sense).  So the cosmic expansion, slowed down by 
gravitational attraction from the matter contents, can instead be 
accelerated if there is sufficient negative pressure material -- dark 
energy.  One can even motivate the Friedmann acceleration equation 
from Newton's second law, including Einstein's admonition that we 
must account for pressure as a form of energy, and hence mass, as well. 

Negative pressure can be made more palatable by pointing out that 
springs and rubber bands essentially have negative pressure.  From the 
first law of thermodynamics we see that the (adiabatic) change in energy 
of a system when the volume changes is given by the negative of the 
pressure.  The hot air inside your body cools when it expands out through 
pursed lips, as a student can easily demonstrate for themselves by blowing 
on their fingers (working as temperature sensors) with lips pursed or 
with mouth open.  Hence the pressure here is a positive quantity.  However, 
when a spring or rubber band is expanded, the energy goes up, implying 
the pressure is a negative quantity.  (Again, students can demonstrate 
this by touching a rapidly stretched rubber band to their 
temperature-sensitive lips.)  

Now perhaps the students have heard that quantum physics predicts that 
energy fields, even the zeropoint energy of the vacuum, should act as 
springs (harmonic oscillators).  Lest their eyes glaze over at the 
mention of field theory, simply explain that a field in physics is 
not so different than a field of grass.  On a lawn or in a meadow, each 
point has a stalk of grass of some height, like an array of springs of 
some length.  A single number at every point 
is simply a scalar field $\phi(x)$ (and a trampled field of grass, or crop 
circles, where each stalk has a length and direction is like a vector field). 
Quantum physics forbids the springs to be absolutely still, they must 
have some fluctuations, stretching and compressing.  But as we saw 
with the rubber band, this leads to a negative pressure, so the vacuum 
itself can have negative pressure. 

Putting it all together, 

\begin{itemize} 

\item Newton: gravitational force depends on acting gravitational mass 

\item Einstein: gravitational mass is equivalent to energy, including 
the pressure 

\item $\quad$ {\bf $\Rightarrow$} Gravitational ``repulsion'' or acceleration 
of the expanding universe can occur if there is sufficient negative pressure 
substance. 

\item Thermodynamics: negative pressure material is allowable, e.g.\ springs 

\item Quantum physics: fields have zeropoint energy and this very vacuum 
acts like springs 

\item $\quad${\bf $\Rightarrow$} Space itself has a vacuum energy with 
negative pressure that could accelerate the expansion of the universe. 

\end{itemize} 

A question than then naturally arises, of considerable pedagogical and 
scientific interest, is what is the fate of our universe.

\section{Fate of the Universe \label{sec:fate}} 

For many years in cosmology, it was declared that geometry equals 
fate.  Models were characterized into open, closed, and critical 
universes (eternally expanding, eventually contracting, and 
asymptotically static), taken to be equivalent to negative, positive, 
and zero spatial curvature models.  For accelerating universes, this 
identification breaks down and one can have open universes with positive 
curvature, for example.  Both the amount and the nature of components 
enter into determining the fate of the universe.  Thus the Newtonian 
picture of throwing a ball into the air and based on the initial 
velocity it will either return to 
earth, orbit, or escape no longer holds; with sufficiently negative 
equation of state (pressure to energy density ratio) the gravitating 
mass of the ball can be negative and a softly thrown ball can 
accelerate away. 

In an accelerating universe, the particle horizon (the region within 
the causally reachable universe) grows rapidly, more rapidly than 
the speed of light, so as time goes on we see an increasingly small 
fraction of the universe; effectively, the limits to our astronomy 
close in as galaxies that used to be visible get pulled away and 
fade out of sight. 
(See Ref.~\ref{tamara}, \ref{horizon} for more on horizons.) 
Matter and radiation get increasingly dilute and the temperature of 
the background radiation tends toward absolute zero.  Whether life 
can survive as the universe ages is a matter of debate (see 
Refs.~\ref{dyson}-\ref{eschat}). 

If the equation of state is even more negative, beyond the cosmological 
constant value of pressure equal and opposite of the energy density, the 
dark energy becomes ``phantom'' (see Ref.~\ref{phantom}).  This leads to 
the energy density of the substance {\it increasing\/} as the expansion 
dilutes it, and the expansion superaccelerating, in a runaway process 
called the Big Rip.  Naively this would lead to the rapid expansion 
eventually tearing apart galaxies, stars, and even atoms (see 
Ref.~\ref{bigrip}) but quantum particle creation might intervene, 
ending the runaway. 

Other scenarios include dark energy fields that roll down their potential 
energy to negative values -- these can lead to a collapsing universe 
as the negative density has a negative gravitational repulsion, i.e.\ 
an attraction again (see Ref.~\ref{doomsday}), related models with 
``sudden singularities'' (e.g.\ Refs.~\ref{sudden1}, \ref{sudden2}), 
and dark energy that waxes and wanes so cosmic acceleration is merely 
episodic (e.g.\ the stochastic model of Ref.~\ref{stochastic}).  Until 
we understand the physical nature of dark energy we will not know the 
fate of our universe.

\begin{enumerate} 

\setcounter{enumi}{\thetempref}

\item \label{tamara} 
``Misconceptions about the Big Bang,'' C.H.\ Lineweaver \& T.M.\ Davis, 
Sci.\ Am.\ {\bf 292}, 36-45 (Mar.\ 2005).  Leads the reader by the hand 
through 
some of the conundra and properties of cosmic expansion and horizons. (E)

\item \label{horizon} 
``Lost horizons,'' G.F.R.\ Ellis \& T.\ Rothman, Am.\ J.\ Phys.\ {\bf 61}, 
883-893 (Oct.\ 1993). Mathematical discussion of various types of horizons 
and related phenomena, and their implications, with abundant diagrams. (I)

\item \label{dyson} 
``Time Without End: Physics and Biology in an Open Universe,'' F.J.\ Dyson, 
Rev.\ Mod.\ Phys.\ {\bf 51}, 447-60 (1979).  Thoughtful and quantitative 
discussion of physical processes and life in an eternally expanding universe 
from a thermodynamic perspective. (I)

\item \label{starkmankrauss} 
``Life, the Universe, and Nothing: Life and Death in an Ever-expanding 
Universe,'' L.M.\ Krauss \& G.D.\ Starkmann, ApJ {\bf 531}, 22-30 (2000). 
Consideration of astronomical observations and life in a cosmological 
constant universe using information theoretic arguments.  Some conclusions 
disagree with Ref.~\ref{dyson} and led to lively debate. (I)

\item \label{eschat}
``Resource Letter: PEs-1: Physical Eschatology,'' M.M.\ {\'C}irkovi{\'c}, 
Am.\ J.\ Phys.\ {\bf 71}, 122-133. Wide ranging and thought provoking 
philosophical and scientific investigation of the future of the universe 
and objects and life in it. (I)

\item \label{phantom} 
``A phantom menace? Cosmological consequences of a dark energy component with 
super-negative equation of state,'' R.R.\ Caldwell, Phys.\ Lett.\ B {\bf 545}, 
23-29 (2002). Influential article arguing that 
motivations exist for considering equations of state more negative than 
the cosmological constant, and examining effects on cosmological 
observations. (A)

\item \label{bigrip} 
``Phantom Energy: Dark Energy with $w<-1$ Causes a Cosmic Doomsday,'' 
R.R.\ Caldwell, M.\ Kamionkowski, \& N.N.\ Weinberg, Phys.\ Rev.\ Lett.\ 
{\bf 91}, 071301 (2003).  Discussion of the dramatic implications of 
phantom energy for the fate of the universe and objects therein, known 
as the Big Rip. (A)

\item \label{doomsday} 
``Observational bounds on cosmic doomsday,'' R.\ Kallosh, J.\ Kratochvil, 
A.\ Linde, E.V.\ Linder, M.\ Shmakova, JCAP {\bf 0310}, 015 (2003).  
Investigation of the lifetime of the universe until the accelerated 
expansion ends and it in fact collapses, if 
dark energy has a simple linear potential.  Gives quantitative 
limits on the time left based on current data, and future projections. (A)

\item \label{sudden1}  
``Sudden future singularities,'' J.D.\ Barrow, Class.\ Quant.\ Grav.\ 
{\bf 21}, L79-L82 (2004).  Discussion of conditions under which 
singularities can occur at a finite time in the future of the universe. (A)

\item \label{sudden2} 
``Necessary and sufficient conditions for big bangs, bounces, crunches, 
rips, sudden singularities and extremality events,'' C.\ Catto{\"e}n \& 
M.\ Visser, Class.\ Quant.\ Grav.\ {\bf 22}, 4913-4930 (2005). Survey 
of several categories of singularities and similar phenomena in the 
properties of the universe. (A)

\item \label{stochastic} 
``Solving the Coincidence Problem: Tracking Oscillating Energy,'' 
S.\ Dodelson, M.\ Kaplinghat, \& E.\ Stewart, Phys.\ Rev.\ Lett.\ 
{\bf 85}, 5276-5279 (2000).  Examines the possibility of cosmic 
acceleration as a periodic or stochastic phenomenon, its amelioration of 
the cosmic coincidence problem, and its observational implications. (A) 

\setcounter{tempref}{\theenumi}

\end{enumerate}

\section{Review Articles \label{sec:review}}

Review articles of dark energy cosmology as a whole or specific topics 
within the field provide an invaluable resource for people interested 
in getting exposed to the concepts, results, progress, and prospects of 
the field.  The categories give a general indication of the main focus 
of the articles and the technical levels provide a rough guide 
to the intended audience. 

\vspace{0.1in}

\noindent{\bf Dark Energy Characteristics} 

\begin{enumerate} 

\setcounter{enumi}{\thetempref}

\item \label{bctrodden} 
``Insights into Dark Energy: Interplay between Theory and Observation,'' 
R.\ Bean, S.\ Carroll, \& M.\ Trodden, arXiv:astro-ph/0510059. (A) 

\item \label{caldwell00}
``An Introduction to Quintessence,'' R.R.\ Caldwell, in {\bf Sources and 
Detection of Dark Matter and Dark Energy in the Universe}, Fourth 
International Symposium, edited by D.B.\ Cline (Springer Verlag, New York, 
2001), pp.\ 74-91. (I)

\item \label{livrellam}
``The Cosmological Constant,'' S.M.\ Carroll, Living Rev.\ Relativity {\bf 4}, 
1 (2001). (A)

\item \label{joyce01}
``Aspects of Cosmology with Scalar Fields,'' M.\ Joyce, habilitation 
thesis 2000, \url{http://supernovae.in2p3.fr/~joyce}. (A)

\item \label{grg}
``The Dynamics of Quintessence, the Quintessence of Dynamics,'' E.V.\ 
Linder, Gen.\ Rel.\ Gravitation in press (2007). (A)

\item \label{padmarev} 
``Darker Side of the Universe,'' T.\ Padmanabhan, in {\bf Proc. 29th Int.\ 
Cosmic Ray Conf.\ 10}, 47-62 (2005). (A) 
%? http://arxiv.org/abs/astro-ph/0510492
% equivalent to http://arxiv.org/abs/astro-ph/0602117 
%Advanced Topics in Cosmology: A Pedagogical Introduction [use first one]

\item \label{peebratra}
``The Cosmological Constant and Dark Energy,'' P.J.E.\ Peebles \& B.\ Ratra, 
Rev.\ Mod.\ Phys.\ {\bf 75}, 559-606 (2003). (A)

\item \label{sahni} 
``Dark Matter and Dark Energy,'' V.\ Sahni, in {\bf The Physics of the 
Early Universe}, edited by E. Papantonopoulos, Lect.\ Notes Phys.\ {\bf 653}, 
141-180 (Springer, Berlin, 2004). (I)
% http://arxiv.org/abs/astro-ph/0403324 

\item \label{sahnstar} 
``Reconstructing Dark Energy,'' V.\ Sahni \& A.\ Starobinsky, Int.\ J.\ 
Mod.\ Phys.\ D {\bf 15}, 2105-2132 (2006). (A)
% http://arxiv.org/abs/astro-ph/0610026

\item \label{straumann} 
``Dark Energy: Recent Developments,'' N.\ Straumann, Mod.\ Phys.\ Lett.\ 
A {\bf 21}, 1083-1098 (2006). (I)
%? http://arxiv.org/abs/hep-ph/0604231 

\item \label{turhut}
``Cosmic Acceleration, Dark Energy and Fundamental Physics,'' M.S.\ Turner 
\& D.\ Huterer, J.\ Phys.\ Soc.\ Japan in press (2007). (I) 

\setcounter{tempref}{\theenumi}

\end{enumerate} 

\vspace{0.1in}

\noindent{\bf Dark Energy and High Energy Physics} 

\begin{enumerate} 

\setcounter{enumi}{\thetempref}

\item \label{bousso} 
``The Holographic Principle,'' R.\ Bousso, Rev.\ Mod.\ Phys.\ {\bf 74}, 
825-874 (2002). (A)

\item \label{copeland}
``Dynamics of Dark Energy,'' E.J. Copeland, M.\ Sami, \& S.\ Tsujikawa, 
Int.\ J.\ Mod.\ Phys.\ D {\bf 15}, 1753-1935 (2006). (A)
% hepth0603057

\item \label{sciamdva} 
``Out of the Darkness,'' G.\ Dvali, Sci.\ Am.\ {\bf 290}, 68-75 
(Feb.\ 2004). (E)

\item \label{string}
``The Issue of Dark Energy in String Theory,'' N.E.\ Mavromatos, 
arXiv:hep-th/0607006. (A)

\item \label{nobbenhuis}
``Categorizing Different Approaches to the Cosmological Constant Problem,'' 
S.\ Nobbenhuis, Found.\ Phys.\ {\bf 36}, 613-680. (A)

\item \label{padma03}
``Cosmological Constant -- the Weight of the Vacuum,'' T.\ Padmanabhan, 
Phys.\ Rept.\ {\bf 380}, 235-320 (2003). (A)

\item \label{pilar}
``Dark energy, gravitation and supernovae,'' P.\ Ruiz-Lapuente, Class.\ 
Quantum Grav.\ {\bf 24}, R91-R111 (2007). (A)

\item \label{straumann2} 
``On the Cosmological Constant Problems and the Astronomical Evidence 
for a Homogeneous Energy Density with Negative Pressure,'' N.\ Straumann, 
in {\bf Vacuum Energy -- Renormalization}, edited by B.\ Duplantier \& 
V.\ Rivasseau (Birkh{\"a}user Verlag, Berlin, 2003), pp.\ 7-52. (A)
%? http://arxiv.org/abs/astro-ph/0203330 
% beginning, e.g. S1-4

\setcounter{tempref}{\theenumi}

\end{enumerate} 

\vspace{0.1in}

\noindent{\bf Dark Energy and Observations} 

\begin{enumerate} 

\setcounter{enumi}{\thetempref}

\item \label{araaadel} 
``Tests of the Gravitational Inverse-Square Law,'' E.G.\ Adelberger, B.R.\ 
Heckel, and A.E.\ Nelson, Ann.\ Rev.\ Nucl.\ Part.\ Sci.\ {\bf 53}, 
77-121 (2003). (A)

\item \label{lecnotes646} 
{\bf The Early Universe and Observational Cosmology}, Lect.\ Notes Phys. 
{\bf 646}, edited by N.\ Breton, 
J.L.\ Cervantes-Cota, and M.\ Salgado (Springer Verlag, Berlin, 2004). (A)

\item \label{sciamcon} 
``The Universe's Invisible Hand,'' C.J.\ Conselice, Sci.\ Am.\ {\bf 296}, 
35-41 (2007). (E)

\item \label{araahudod} 
``Cosmic Microwave Background Anisotropies,'' W.\ Hu \& S.\ Dodelson, 
ARA\&A {\bf 40}, 171-216 (2002). (A)

\item \label{kirshner}
``Throwing Light on Dark Energy,'' R.P.\ Kirshner, Science {\bf 300}, 
1914-1918 (2003). (I)

\item \label{livrelstr} 
``Measuring Our Universe from Galaxy Redshift Surveys,'' O.\ Lahav \& Y.\ 
Suto, Living Rev.\ Relativity {\bf 7}, 8 (2004). (A) 

\item \label{araaleib} 
``Cosmological Implications from Observations of Type Ia Supernovae,'' 
B.\ Leibundgut, ARA\&A {\bf 39}, 67-98 (2001). (A)

\item \label{cern}
``On the Trail of Dark Energy,'' E.V.\ Linder, CERN Courier {\bf 43} (7), 
23-25 (Jul.\ 2003). (I) 

\item \label{ptodayperlsch} 
``Supernovae, Dark Energy, and the Accelerating Universe,'' S.\ Perlmutter, 
Phys.\ Today {\bf 56}, 53-60 (Apr.\ 2003). (E)

\item \label{perlschm} 
``Measuring Cosmology with Supernovae,'' S.\ Perlmutter \& B.P.\ Schmidt, 
in {\bf Supernovae and Gamma Ray Bursters}, edited by K.\ Weiler, Lect.\ 
Notes Phys.\ {\bf 598}, 195-217 (Springer Verlag, Berlin, 2003). (I)
% http://arxiv.org/abs/astro-ph/0303428 

%Stellar Candles for the Extragalactic Distance Scale
%Series: Lecture Notes in Physics , Vol. 635 
%Alloin, Danielle; Gieren, Wolfgang (Eds.) 2003

\item \label{araarefr} 
``Weak Gravitational Lensing by Large-Scale Structure,'' A.\ Refregier, 
ARA\&A {\bf 41}, 645-669 (2003). (A)

\item \label{dhw}
``Dark Energy: The Observational Challenge,'' D.H.\ Weinberg, New Astron.\ 
Rev.\ 49, 337-345 (2005). (I)
% http://arxiv.org/abs/astro-ph/0510196

\setcounter{tempref}{\theenumi}

\end{enumerate} 

\vspace{0.1in}

\noindent{\bf Dark Energy and Cosmological Parameters} 

\begin{enumerate} 

\setcounter{enumi}{\thetempref}

\item \label{snowmass}
``Cosmological Parameters, Dark Energy and Large Scale Structure,'' S.E.\ 
Deustua, R.R.\ Caldwell, P.\ Garnavich, L.\ Hui, A.\ Refregier, 
arXiv:astro-ph/0207293. (I) 

\item \label{teachkrauss} 
``The History and Fate of the Universe,'' L.M.\ Krauss, Phys.\ Teacher 
{\bf 41}, 146-155 (2003). (E)

\item \label{pdglahlid} 
``The Cosmological Parameters 2006,'' O. Lahav \& A.R.\ Liddle, in 
{\bf The Review of Particle Physics 2006}, edited by W.-M.\ Yao et al., 
J.\ Phys.\ G {\bf 33}, 224-232 (2006). (I) 

\item \label{sciamoststein} 
``The Quintessential Universe,'' J.P.\ Ostriker \& P.J.\ Steinhardt, 
Sci.\ Am.\ {\bf 284}, 46-53 (Jan.\ 2001). (E)

\item \label{sciamriesstur} 
``From Slowdown to Speedup,'' A.G.\ Riess \& M.S.\ Turner, Sci.\ Am.\ 
{\bf 290}, 62-67 (Feb.\ 2004). (E)

\item \label{wmap3}
``Three-Year Wilkinson Microwave Anisotropy Probe (WMAP) Observations: 
Implications for Cosmology,'' D.N.\ Spergel et al., Ap.\ J.\ 
%{\bf ???}, ??-?? 
in press (2007). (A)

\item \label{tegmark}
``Measuring spacetime: from big bang to black holes,'' M.\ Tegmark, 
Science {\bf 296}, 1427-1433 (2002). (I) 
%in Ref.~\ref{lecnotes646}, Lect.\ Notes Phys.\ {\bf 646}, 169-189 (2004). (I)
% http://arxiv.org/abs/astro-ph/0207199

\setcounter{tempref}{\theenumi}

\end{enumerate}

\section{Other Resources \label{sec:other}} 

\subsection{Books \label{sec:book}}

Perhaps because dark energy and the accelerating universe is such 
a rapidly developing subject and a mystery regarding its physical 
origin and properties, there is as yet no definitive book on the 
popular or monograph levels.  Since the field is such a moving target, 
those seeking less technical approaches to the subject should look 
to the general journal articles cited under Review Articles and to 
the websites listed below.  For more advanced presentations, consult 
the technical journal articles under Review Articles and the conference 
proceedings and websites listed below. 

For piecemeal but still useful discussions of aspects of the accelerating 
expansion, in particular cosmological probes and components with 
general equations of state, one can explore sections of 
Refs.~\ref{bergstrom}-\ref{ryden}.  A compendium of the state of 
probing dark energy at the time of the Snowmass 2001 meeting on the 
``Future of Particle Physics'' is given in Ref.~\ref{rbde}.

\begin{enumerate} 

\setcounter{enumi}{\thetempref}

\item \label{bergstrom} 
{\bf Cosmology and Particle Astrophysics}, L.\ Bergstrom \& A.\ Goobar 
(Springer-Verlag, Berlin, 2004 2nd ed.).  Contains a brief but 
informative subsection on dark energy and scalar fields, and otherwise 
gives a good and wide ranging overview of cosmology, relativity, and 
particle physics.  Written on the beginning to mid graduate level. (A)

\item \label{dodelson} 
{\bf Modern Cosmology}, S.\ Dodelson (Academic Press, San Diego, 2003). 
Contains a couple of small subsections on dark energy, but is rich 
in general cosmological concepts and applications.  Written on the 
graduate level and as a professional reference. (A)

\item \label{fpoc} 
{\bf First Principles of Cosmology}, E.V.\ Linder (Addison-Wesley, London, 
1997).  Contains several discussions about dark energy (before it got 
the name), and light propagation and cosmological probes in generalized 
cosmologies.  Dated on the observational side but gives care to physical 
analogies and concepts on the theory side.  Written on the upper 
undergraduate to beginning graduate level. (I)

\item \label{ryden} 
{\bf Introduction to Cosmology}, B.\ Ryden (Benjamin Cummings, 2002). 
Contains a couple of brief mentions of dark energy, but provides a clear 
general overview of cosmology, with background concepts and observations. 
Written on the upper undergraduate to beginning graduate level. (I)

\item \label{rbde} 
{\bf Resource Book on Dark Energy}, edited by E.V.\ Linder (Berkeley Lab, 
Berkeley, 2001).  Collection of some 40 articles on dark energy by 100 
authors, giving a many angled perspective on the subject at the time of 
the 2001 Snowmass Symposium on the Future of Particle Physics.  Also 
online at \url{http://supernova.lbl.gov/~evlinder/sci.html}. (I,A)

\setcounter{tempref}{\theenumi}

\end{enumerate}

\subsection{Conferences \label{sec:conference}}

Conference proceedings and talks posted on conference websites provide 
valuable snapshots of research in progress, as well as sometimes overview 
talks.  Here we list some of the main conference series with 
sessions concentrating on dark energy.

\begin{enumerate} 

\setcounter{enumi}{\thetempref}

\item \label{cosmo}
Cosmo 07 \url{http://www.cosmo07.info}.  Annual international conference 
since 1997, covering inflation, the early universe, dark energy, and 
other topics.  Talks are posted online. (A)

\item \label{heid}
Dark 2007 \url{http://www.physics.usyd.edu.au/dark2007}.  Roughly biennial 
international conference since 1996, also known as Heidelberg International 
Conference on Dark Matter in Astro and Particle Physics, covering dark 
matter, astroparticle physics, large scale structure, and dark energy.  
Talks are not always posted online but conference proceedings are 
published, e.g.\ by Springer. (A)

\item \label{dsu}
Dark Side of the Universe \url{http://www.ftpi.umn.edu/dsu07}.  Annual 
international workshop since 2005, treating both theory and experiment related 
to dark matter and dark energy.  Talks are posted online and conference 
proceedings sometimes published, e.g.\ in AIP Conf.\ Proc. (A)

\item \label{sfe} 
Santa Fe Cosmology Summer Workshop 
\url{http://t8web.lanl.gov/people/salman/sf07}.  Annual meeting since 
1999 that combines an interactive summer school for students and postdocs 
with a workshop for researchers.  Very useful for people starting out in 
the field.  Concentrates on large scale structure, CMB, dark matter, and 
dark energy.  Talks are sometimes posted online.  (A,I)

\item \label{ucla}
Sources and Detection of Dark Matter and Dark Energy in the Universe 
\url{http://www.physics.ucla.edu/hep/dm06/dm06.htm}.  Biennial conference 
since 1994, also known as the UCLA Symposium, covering dark matter, 
cosmology, and dark energy.  Talks are posted online and conference
proceedings published, e.g.\ in Nucl.\ Ph.\ B. (A)

\item \label{texas}
Texas Symposium on Relativistic Astrophysics \url{http://www.texas06.com}. 
Biennial international conference since 1963, concentrating on relativistic 
astrophysics, with lesser coverage of cosmology and dark energy.  Talks 
are generally posted online and conference proceedings generally 
published. (A)

\setcounter{tempref}{\theenumi}

\end{enumerate}

\subsection{Discussion Boards \label{sec:disc}}

Online community discussion forums provide a more informal view of 
questions and developments in scientific research.  Two forums run by 
professional physicists include Refs.~\ref{cosmocoffee}, \ref{cosmicvariance}. 

\begin{enumerate} 

\setcounter{enumi}{\thetempref}

\item \label{cosmocoffee} 
CosmoCoffee \url{http://cosmocoffee.info}.  Online interactive discussion 
forum containing sections for comments on arXiv.org papers, technical 
issues in cosmology analysis, and general audience cosmology questions. (A,I)

\item \label{cosmicvariance} 
Cosmic Variance \url{http://cosmicvariance.com}. 
Blog posts from highly regarded physicist contributors (not necessarily 
on technical physics topics, but there is a general dark energy and 
particle physics theme).  Also open comment forum on the posts. (I)

\setcounter{tempref}{\theenumi}

\end{enumerate}

\subsection{Education Websites \label{sec:edu}}

Relatively few websites exist providing broad educational resources 
in dark energy cosmology.  Since websites rapidly come and go, or 
cease being updated, a focused search on the internet may provide 
the most useful information.  Current sites with a strong concentration 
on dark energy include Ref.~\ref{univadv}, \ref{scires}; unfortunately,  
ongoing and next generation observational programs provide few practical, 
general educational resources about dark energy, but see Ref.~\ref{wmapedu}, 
\ref{snapepo}.

\begin{enumerate} 

\setcounter{enumi}{\thetempref}

\item \label{univadv} 
Universe Adventure \url{http://universeadventure.org}.  Linked series 
of webpages teaching aspects of cosmology and the history and fate of 
the universe, aimed at a high school and general audience. 
Contains abundant graphics, mini-quizzes, glossary, and teacher resources. 
Similar to the award-winning ParticleAdventure.org, and extended by 
``History and Fate of the Universe'' charts (see 
\url{http://pdg.lbl.gov/fate-history/posters.html}) which 
Ref.~\ref{teachkrauss} was designed to accompany.  (E)

\item \label{scires} 
Cosmology Resources 
\url{http://supernova.lbl.gov/~evlinder/scires.html}.  A grab bag of 
cosmology resources ranging from additions to Ref.~\ref{fpoc}, summer 
school lectures, faqs, and basic formulas, to dark energy humor.  Levels 
vary from public audience talks to technical papers. (E,I,A) 

\item \label{hiddendim}
Hidden Dimensions \url{http://hiddendimensions.org}.  Discusses the 
role of extra dimensions in various puzzles including particle physics, 
gravity, and the accelerating universe.  Suitable for classroom use 
or individual exploration, contains several animations. (E)

\item \label{wmapedu}
WMAP Cosmology 101 \url{http://wmap.gsfc.nasa.gov/m_uni.html}.  Tutorial 
from the Wilkinson Microwave Anisotropy Probe mission on cosmology with 
an emphasis on the cosmic microwave background, but contains some mention 
of dark energy.  Includes glossary and teacher resources. (E) 

\item \label{snapepo}
%Dark Energy Education Outreach \url{http://snap.lbl.gov/EPO}. 
Dark Energy Education Outreach \url{http://snap.lbl.gov/education}. 
Linked series of webpages introducing many aspects of 
dark energy and the accelerating universe, aimed at students and 
general audience.  Contains graphics and glossary.  While originated 
for the Supernova/Acceleration Probe (SNAP) project, much of the material 
is general, giving good summaries of the puzzles facing scientists trying 
to understand our universe. (E)

\setcounter{tempref}{\theenumi}

\end{enumerate}

\acknowledgments

I gratefully acknowledge use of the NASA ADS abstract service, and thank 
Robert Caldwell for some suggested references. 
This work has been supported in part by the Director, Office of Science,
Department of Energy under grant DE-AC02-05CH11231.

\end{document}